\documentclass{elsarticle}

\usepackage[utf8]{inputenc}
\usepackage{amsmath}
\usepackage[mathscr]{euscript}
\usepackage{gensymb,amsfonts}
\usepackage{amssymb}
\usepackage{array}
\usepackage[T1]{fontenc}
\usepackage[margin=1in]{geometry}
\usepackage{lmodern}
\usepackage{tabularx}
\usepackage{fancyhdr}
\usepackage{graphicx}
\graphicspath{{images/}}
\usepackage[english]{babel}
\usepackage{amsthm,mathtools}
\usepackage{braket}
\usepackage{tikz}
\usepackage{amsbsy}
\usepackage{mathrsfs}
\usepackage{epsfig}
\usepackage{array}
\usepackage{textcomp}
\usepackage{color}
\usepackage{bm,hyperref}
\usepackage{tensor} % https://tex.stackexchange.com/a/25976
\usepackage[titletoc,toc,title]{appendix}
\usepackage{ulem}

\newcommand{\beq}{\begin{equation}}
\newcommand{\eeq}{\end{equation}}

\newcommand{\refcite}[1]{Ref.~\cite{#1}}
\newcommand{\eqnref}[1]{Eq.~(\ref{#1})}
\newcommand{\eqsref}[1]{Eqs.~(\ref{#1})}
\newcommand{\figref}[1]{Fig.~\ref{#1}}
\newcommand{\sfigref}[2]{Fig.~\hyperref[#1]{\ref{#1}(#2)}}
\newcommand{\tabref}[1]{Tab.~\ref{#1}}
\newcommand{\secref}[1]{Sec.~\ref{#1}}
\newcommand{\appref}[1]{Appendix~\ref{#1}}

\newcommand{\dd}{\mathrm{d}}
\newcommand{\vx}{\mathbf{x}}

\hypersetup{
  colorlinks = true,
  urlcolor = blue,
  linkcolor = black,
  pdfauthor = {Kevin Slagle, Abhinav Prem, Michael Pretko},
  pdftitle = {Slagle et al. - 2018 - Symmetric Tensor Gauge Theories on Curved Spaces}
}

\biboptions{numbers,sort&compress} % use citation ranges: https://tex.stackexchange.com/a/82294

\begin{document}

\begin{frontmatter}

\title{Symmetric Tensor Gauge Theories on Curved Spaces}

\author{Kevin Slagle}
\address{Department of Physics, University of Toronto \\ Toronto, Ontario M5S 1A7, Canada}
\address{Department of Physics and Institute for Quantum Information and Matter \\ California Institute of Technology, Pasadena, California 91125, USA}
\ead{kslagle@caltech.edu}

\author{Abhinav Prem}

\author{Michael Pretko}
\address{Department of Physics and Center for Theory of Quantum Matter \\ University of Colorado, Boulder, CO 80309}

\begin{abstract}
Fractons and other subdimensional particles are an exotic class of emergent quasi-particle excitations with severely restricted mobility. A wide class of models featuring these quasi-particles have a natural description in the language of symmetric tensor gauge theories, which feature conservation laws restricting the motion of particles to lower-dimensional sub-spaces, such as lines or points. In this work, we investigate the fate of symmetric tensor gauge theories in the presence of spatial curvature. We find that weak curvature can induce small (exponentially suppressed) violations on the mobility restrictions of charges, leaving a sense of asymptotic fractonic/sub-dimensional behavior on generic manifolds. Nevertheless, we show that certain symmetric tensor gauge theories maintain sharp mobility restrictions and gauge invariance on certain special curved spaces, such as Einstein manifolds or spaces of constant curvature.
\end{abstract}

\end{frontmatter}

{\hypersetup{linkcolor=black} % make table of contents links black: https://tex.stackexchange.com/a/88416
\tableofcontents}

%%%%%%%%%%%%%%%%%%%%%%%%%%%%%%%%%%%%%%%%

\section{Introduction}
\label{intro}

Gauge theories are ubiquitous in modern theoretical physics, across numerous disciplines. In high energy physics, gauge fields form a fundamental component of the Standard Model. In condensed matter physics, gauge theories also provide a natural theoretical description of the phenomenon of fractionalization, as seen in the context of fractional quantum Hall systems, spin liquids, and other topological phases of matter~\cite{Laughlin,wenniu,mooreread,wen2002,balents2002,moessner,levinwen,sondhi,lucile,moroz}. In all of these situations, the most commonly encountered gauge theories involve a gauge field that transforms as a vector under spatial symmetries, in close analogy with the vector potential $\vec{A}$ of Maxwell theory. In principle, however, a gauge field can transform like an arbitrary tensor object, which leads one to consider theories with higher-rank tensor gauge fields. Certain theories of this type are well-studied, such as the higher-form (anti-symmetric tensor) gauge theories, commonly encountered in string theory and in three-dimensional topological phases~\cite{kalb,kapustin}. It seems natural to also consider the symmetric tensors characteristic of higher-spin gauge fields. In the context of high energy physics, however, Lorentz invariance severely limits the set of possible theories we can consider.  Various difficulties associated with preserving Lorentz invariance prevent the formulation of generic interacting theories of symmetric tensor gauge fields, leaving only a few consistent possibilities, such as Einstein gravity and Vasiliev theory~\cite{vasiliev}.

In the context of condensed matter theory, however, an emergent gauge field need not exhibit Lorentz invariance, thereby vastly increasing the set of possible theories we can consider. Nevertheless, symmetric tensor gauge theories have until recently received only limited attention in the condensed matter literature~\cite{cenke,wengu,horava,rasmussen}, since it was unclear whether such theories held any qualitatively new phenomena. Within the past few years, however, it has been realized that symmetric tensor gauge theories can host a new class of emergent quasi-particles with unusual restrictions on their mobility~\cite{sub,genem,higgs1,higgs2}. The most famous quasi-particle of this type is the fracton excitation, which is strictly immobile in isolation, but can often move in bound states with other fractons~\footnote{The notable exceptions are the type-II fracton models, such as Haah's code~\cite{haah}, in which there are no non-trivial mobile bound states.}.  More generally, there exist subdimensional particles restricted to motion within lower-dimensional subspaces, such as lines or planes, within a three-dimensional space. Fractons and other subdimensional particles have been the subject of intense recent study~\cite{chamon,castelnovo,bravyi,haah,haah2,yoshida,fracton1,fracton2,sub,genem,williamson,DevakulWilliamson,prem,han,sagar,mach,hsieh,slagle1,screening,nonabelian,decipher,balents,slagle2,chiral,prem2,regnault,valbert,devakul,regnault2,han2,albert,leomichael,gromov,shirley,foliatedEntanglement,foliatedExcitations,shirleyCheckerboard,shirleyGauging,slagle3,pai,yizhi1,cagenet,twisted,higgs1,higgs2,deconfined,ungauging,fractalsym,symfrac,BulmashFractal,pinchpoints}, and the connection with symmetric tensor gauge theories has provided links with numerous other areas of physics, such as elasticity theory~\cite{leomichael,pai,gromov}, gravity~\cite{mach}, deconfined quantum criticality~\cite{deconfined}, and quantum Hall physics~\cite{chiral,prem2}. An excellent review describing recent progress in the field can be found in \refcite{fractonreview}. 

While symmetric tensor gauge theories have been the subject of intense recent attention, there remain important gaps in our understanding of these theories.  Perhaps most notably, very little is known about how these theories respond to the presence of spatial curvature. Previous studies of gapped fracton models have found a curious sensitivity to the geometry of the underlying lattice, which can affect not only the ground state degeneracy but also the mobility of excitations~\cite{slagle2,slagle3,shirley}.  We expect similar considerations to carry over to the tensor gauge theories, where difficulties in introducing curvature have previously been encountered~\cite{gromov}. It is not clear {\it a priori} whether the mobility restrictions seen in flat space remain intact once curvature is introduced. Furthermore, it is not immediately obvious how the notion of restriction to a line or plane generalizes to curved spaces.

In this work, we fill in this gap by investigating the fate of the symmetric tensor gauge theories in the presence of spatial curvature.  In \secref{sec:curvature}, we argue that a generic theory of this type will suffer from curvature-induced violations of its mobility restrictions and conservation laws on a generic manifold, with a corresponding loss of gauge invariance.  However, we argue that the effective inertial mass of a particle in its forbidden directions grows exponentially in the inverse curvature as the flat space limit is approached [\eqnref{eq:hop}].  The result is that a large amount of curvature (relative to the scale of the energy gap of a fracton dipole) is required to observe significant mobility in restricted directions, while weak curvature has a negligible effect on the properties of the system. For small amounts of curvature, it may take longer than the age of the universe to observe significant motion in the restricted directions.  In this sense, we regard symmetric tensor gauge theories in weakly curved space as hosting asymptotic mobility restrictions, in close analogy with asymptotic many-body localization~\cite{asymptotic1,asymptotic2}, where thermalization occurs only at exponentially long times.

While generic symmetric tensor gauge theories can exhibit weak violations of mobility restrictions on generic manifolds, in \secref{sec:fractons on einstein} we show that there exists a subset of these theories which are more robust against the introduction of curvature.  We find that certain tensor gauge theories, such as the three-dimensional traceless scalar charge theory, retain sharp mobility restrictions and gauge invariance on Einstein manifolds, which correspond to solutions of Einstein's equation sourced only by a (spatially dependent) cosmological constant.  In other cases, such as the two-dimensional traceless scalar charge theory, gauge invariance is maintained only on spaces of constant curvature.  The fate of the mobility restrictions and gauge symmetry of several other tensor gauge theories is summarized in \tabref{tab:summary}.

Throughout this work, we consider only spatial curvature and not space-time curvature; $i.e.$ we consider spacetime metrics $g_{\mu\nu}$ where $g_{00} = -1$ and $g_{0a} = g_{a0} = 0$.  We use Greek letters to denote spacetime indices ($\mu,\nu=0,1,...,D$) while Latin letters denote spatial indices ($a,b,c,i,j,k=1,...,D$).

%%%%%%%%%%%%%%%%%%%%%%%%%%%%%%%%%%%%%%%%

\section{Review of Symmetric Tensor Gauge Theories}
\label{sec:review}

Before introducing spatial curvature, we will begin by reviewing the symmetric tensor gauge theories discussed in this work on flat manifolds~\cite{sub}.  We focus on rank-two tensor gauge fields for simplicity, though similar analyses apply to tensor gauge theories of arbitrary rank. We will work in the Hamiltonian formalism, where we define a symmetric tensor gauge field $A_{ij}$ and its canonical conjugate $E_{ij}$, which we regard as a generalized electric field tensor. Using these fundamental ingredients, a gauge theory can be constructed by specifying the generalized Gauss's law on the electric tensor, which in turn specifies the gauge transformation.

\subsection{Scalar Charge Theory}

In the scalar charge theory, the defining Gauss's law takes the form
\begin{align}
\partial_i\partial_j E^{ij} = \rho \label{eq:scalar gauss},
\end{align}
for a scalar charge density $\rho$.  Within the charge-free sector, the system is invariant under the following gauge transformation:
\begin{equation}
A_{ij}\rightarrow A_{ij} + \partial_i\partial_j\lambda \label{eq:scalar gauge},
\end{equation}
for a scalar function $\lambda$ with arbitrary spatial dependence.  The most general low-energy Hamiltonian consistent with this gauge transformation schematically takes the same form as in Maxwell theory,
\begin{equation}
H = \int \dd^d x\, \frac{1}{2}(E^2 + B^2),
\label{ham}
\end{equation}
where $E^2 = E^{ij}E_{ij}$, and $B$ represents the (lowest-order, $i.e.$ least number of derivatives) gauge-invariant magnetic field operator.
%As emphasized in \refcite{sub}, a $(\partial_i \partial_j E^{ij})^2$ term should also be added to the Hamiltonian if one desires that the phase of the ground state of the Hamiltonian is stable to perturbations that are not gauge invariant.
For example, in three spatial dimensions, we have a (traceless) magnetic tensor of the form
\begin{equation}
  B^{ij} = \delta^{jc} \epsilon^{iab} \partial_a A_{bc} \label{eq:scalarB}
\end{equation}
and $B^2 = B^{ij}B_{ij}$.  In this case, we obtain five gapless gauge modes with linear dispersion, $\omega\sim k$.

While the low-energy Hamiltonian is important, the most interesting aspect of this tensor gauge theory follows directly from the defining Gauss's law. In addition to conservation of charge,
\begin{equation}
\int \dd^d x\,\rho = \textrm{constant},
\end{equation}
this theory also features conservation of dipole moment,
\begin{equation}
\int \dd^d x\,(\rho\vec{x}) = \textrm{constant},
\end{equation}
which follows immediately from integration by parts, making use of the two derivatives in Gauss's law. This extra conservation law has the unusual consequence that an isolated charge is strictly locked in place, since motion of an individual charge would change the total dipole moment of the system. In other words, charges in this tensor gauge theory behave as fractons.  While individual charges cannot move, the dipole conservation law still allows for mobile bound states. Specifically, a dipolar bound state of two equal and opposite charges is a fully mobile (and stable) excitation, which is free to move in any direction provided it preserves the orientation of its dipole moment.

The scalar charge theory can also be described by either of the following Lagrangians
\begin{align}
\begin{split}
  L(E,\phi,A) &= - E^{ij} (\partial_t A_{ij} - \partial_i \partial_j \phi) - \frac{1}{2} E^{ij} E_{ij} - \frac{1}{2} B^{ij} B_{ij} \\
  L(\phi,A) &= \frac{1}{2} (\partial_t A_{ij} - \partial_i \partial_j \phi)^2 - \frac{1}{2} B^{ij} B_{ij} \label{eq:scalarLag}
\end{split}
\end{align}
where $B^{ij}$ is given by \eqnref{eq:scalarB}.
The first Lagrangian involves three fields:
  $E^{ij}$, $\phi$, and $A_{ij}$
  where $E^{ij} = E^{ji}$ and $A^{ij} = A^{ji}$ are symmetric.
The $E^{ij} \partial_t A_{ij}$ term encodes the fact that $A$ and $E$ are conjugate variables.
$\phi$ is a Lagrange multiplier that enforces Gauss's law [\eqnref{eq:scalar gauss}] and transforms as $\phi \to \phi + \partial_t \lambda$ under gauge transformations.
$\frac{1}{2} E^{ij} E_{ij} + \frac{1}{2} B^{ij} B_{ij}$
  is the Hamiltonian [\eqnref{ham}].
The second Lagrangian is obtained by integrating out $E^{ij}$.

\subsection{Traceless Scalar Charge Theory}
\label{sec:traceless scalar}

We may also define a slightly modified theory by taking the scalar charge theory and imposing a local tracelessness constraint,
\begin{equation}
  \delta_{ij} E^{ij} = 0 \label{eq:traceless scalar},
\end{equation}
in addition to the Gauss's law [\eqnref{eq:scalar gauss}]. Imposing this constraint requires the following local symmetry (else time evolution would violate the constraint):
\begin{equation}
  A_{ij} \to A_{ij} + \delta_{ij} \mu , \label{eq:scalar traceless gauge}
\end{equation}
where $\mu(x)$ has spatial dependence.
This local symmetry results in the following traceless and symmetric magnetic field in three spatial dimensions:
\begin{equation}
  B^{ij} = \frac{1}{2} \delta^{jc} \epsilon^{iab} \partial_a A_{bc} + (i \leftrightarrow j) \label{eq:scalar traceless mag}.
\end{equation}

The traceless constraint results in a quadrupole conservation law
\begin{equation}
  \int \dd^d x\, x^2 \rho = 0,
\end{equation}
which makes a dipole a 2D particle, moving only perpendicular to its dipole moment, instead of a fully mobile particle as in the traceful theory. The traceless constraint also results in a duality between the electric and magnetic sectors (similar to the usual vector $U(1)$ gauge theory), which does not exist for the scalar charge theory~\cite{genem}.

The traceless scalar charge theory can also be described by either of the following Lagrangians:
\begin{align}
\begin{split}
  L(E,\phi,\theta,A) &= - E^{ij} (\partial_t A_{ij} - \partial_i \partial_j \phi - \delta_{ij} \theta) - \frac{1}{2} E^{ij} E_{ij} - \frac{1}{2} B^{ij} B_{ij} \\
  L(\phi,\theta,A) &= \frac{1}{2} (\partial_t A_{ij} - \partial_i \partial_j \phi - \delta_{ij} \theta)^2 - \frac{1}{2} B^{ij} B_{ij}
\end{split}
\end{align}
where $B^{ij}$ is given by \eqnref{eq:scalar traceless mag}.
The Lagrangians are similar to \eqnref{eq:scalarLag},
  except a $\theta$ field is added to impose the traceless constraint [\eqnref{eq:traceless scalar}] on $E^{ij}$.
(In the Lagrangian, $E^{ij}$ is not traceless until $\theta$ is integrated out.)
$\theta \to \theta + \partial_t \mu$ under the local symmetry [\eqnref{eq:scalar traceless gauge}].

\subsection{Vector Charge Theory}
\label{vectorreview}

Using the same fundamental variables, we may also define a completely different theory by specifying a different choice of Gauss's law.  In the vector charge theory, the Gauss's law takes the form
\begin{equation}
\partial_iE^{ij} = \rho^j,
\end{equation}
for a vector charge density $\rho^j$.  The corresponding gauge transformation within the charge-free sector is
\begin{equation}
A_{ij}\rightarrow A_{ij} + \frac{1}{2}(\partial_i\lambda_j + \partial_j\lambda_i), \label{eq:vectorGauge}
\end{equation}
for a vector gauge function $\lambda_i$ with arbitrary spatial dependence. The low-energy Hamiltonian for this theory takes the same schematic Maxwell form seen in \eqnref{ham}, but with a different magnetic field operator. For example, in three dimensions, the magnetic tensor is symmetric ($B^{ij} = B^{ji}$) and takes the form
\begin{equation}
  B^{ij} = \epsilon^{iab}\epsilon^{jcd}\partial_a\partial_cA_{bd}, \label{eq:vecB}
\end{equation}
leading to three gapless gauge modes with quadratic dispersion, $\omega\sim k^2$.

As in the scalar charge theory, the most interesting feature of the vector charge theory is its unusual set of conservation laws. This theory exhibits conservation of its vector-valued charge
\begin{equation}
\int \dd^d x\,\vec{\rho} = \textrm{constant},
\end{equation}
and also conservation of the angular moment of charge, which in three dimensions takes the form
\begin{equation}
\int d^3x\,(\vec{x}\times\vec{\rho}) =  \textrm{constant}.
\end{equation}
The consequence of this conservation law is that a charged particle can only move along the line spanned by its charge vector.  In this sense, the charges of this theory behave like one-dimensional particles\footnote{Regarding terminology, a `one-dimensional particle' is a point-like excitation that can only move in a single direction even though it may live in two or more spatial dimensions.}, restricted to motion along a straight line. However, a dipole of two vector charges (i.e. two vector charges with opposite charge displaced in a direction orthogonal to the charge vectors) is fully mobile.

The vector charge theory can also be described by either of the following Lagrangians:
\begin{align}
\begin{split}
  L(E,\phi,A) &= - E^{ij} \big[ \partial_t A_{ij} - \tfrac{1}{2}(\partial_i \phi_j + \partial_j \phi_i) \big] - \frac{1}{2} E^{ij} E_{ij} - \frac{1}{2} B^{ij} B_{ij} \\
  L(\phi,A) &= \frac{1}{2} \big[ \partial_t A_{ij} - \tfrac{1}{2}(\partial_i \phi_j + \partial_j \phi_i) \big]^2 - \frac{1}{2} B^{ij} B_{ij} \label{eq:vecLag}
\end{split}
\end{align}
where $B^{ij}$ is given by \eqnref{eq:vecB}.
The Lagrangians are similar to \eqnref{eq:scalarLag},
  except $\phi_i \to \phi_i + \partial_t \lambda_i$ under gauge transformations.

\subsection{Traceless Vector Charge Theory}
\label{tracelessvectorreview}

As with the scalar charge theory, we can also define a slightly modified vector charge theory by imposing a local tracelessness constraint
\begin{equation}
\delta_{ij}E^{ij}=0 \label{eq:vecTraceless}
\end{equation}
in addition to the defining vector Gauss's law. This constraint once again forces the theory to be invariant under the extra local symmetry
\begin{equation}
  A_{ij}\rightarrow A_{ij}+\delta_{ij} \mu. \label{eq:tracelessVecGauge}
\end{equation}
In this case, the magnetic field tensor takes the form
\begin{equation}
B^{ij} = \frac{1}{2} \epsilon^{jab} \eta^{ik} \eta^{cd} (\partial_a\partial_c\partial_k A_{bd} - \partial_a\partial^2 A_{bk})+(i\leftrightarrow j). \label{eq:tracelessVecB}
\end{equation}
In addition to the two conservation laws of the traceful vector charge theory, the traceless vector charge theory has two further conservation laws,
\begin{align}
\int d^dx\,(\vec{\rho}\cdot\vec{x})&=\textrm{constant}, & \int d^dx\,[(\vec{\rho}\cdot\vec{x})\vec{x} - \frac{1}{2}x^2\vec{\rho}] &= \textrm{constant}.
\end{align}
These extra conservation laws have two notable consequences.  First, the vector charges, which were formerly one-dimensional, now become fully locked in place, $i.e.$ fractons, since longitudinal motion is not consistent with the new conservation laws.  Second, the dipolar bound states, formerly fully mobile, now become one-dimensional particles~\cite{genem}.

The traceless vector charge theory can also be described by either of the following Lagrangians:
\begin{align}
\begin{split}
  L(E,\phi,\theta,A) &= - E^{ij} \big[ \partial_t A_{ij} - \tfrac{1}{2}(\partial_i \phi_j + \partial_j \phi_i) - \delta_{ij} \phi \big] - \frac{1}{2} E^{ij} E_{ij} - \frac{1}{2} B^{ij} B_{ij} \\
  L(\phi,\theta,A) &= \frac{1}{2} \big[ \partial_t A_{ij} - \tfrac{1}{2}(\partial_i \phi_j + \partial_j \phi_i) - \delta_{ij} \phi \big]^2 - \frac{1}{2} B^{ij} B_{ij}
\end{split}
\end{align}
where $B^{ij}$ is given by \eqnref{eq:tracelessVecB}.
The Lagrangians are similar to \eqnref{eq:vecLag},
  except a $\theta$ field is added to impose the traceless constraint [\eqnref{eq:vecTraceless}].
  $\theta \to \theta + \partial_t \mu$ under the local symmetry [\eqnref{eq:tracelessVecGauge}].

%%%%%%%%%%%%%%%%%%%%%%%%%%%%%%%%%%%%%%%%

\section{Curvature-Induced Violation of Mobility Restrictions}
\label{sec:curvature}

The most crucial piece of physics associated with symmetric tensor gauge theories and their mobility restrictions is the presence of higher moment charge conservation laws, such as the conservation of dipole moment, $\int \dd^d x\,(\rho\vec{x})$, which explicitly involves the spatial coordinates of the system. Unfortunately, conservation laws of this form are inherently difficult to extend to curved spaces, where a smooth global coordinate chart need not exist. One can instead try to take a more local perspective on conservation laws, imposing that all local operations must preserve the locally defined dipole moment within some small patch where a coordinate chart can be defined.  Even in this case however, curvature proves fatal to the sharp mobility restrictions seen in flat space.

The curvature-induced breakdown of fractonic and subdimensional behavior can be easily understood in terms of parallel transport of particles carrying the conserved quantity. For concreteness, let us focus on the scalar charge theory, which in flat space features conservation of dipole moment and fractonic behavior of charges.  Even if we impose that the motion of a fracton via local operators creates extra dipoles (thus preserving the local dipole moment), the immobility of fractons can still be violated via the following process:

\begin{enumerate}

\item A fracton moves in a given direction, emitting a dipole oriented in the opposite direction.

\item The mobile dipole is parallel transported around a closed curve, rotating due to non-trivial curvature.

\item The dipole, with its dipole moment now re-oriented, is re-absorbed by the fracton, which ends up in a different position from where it started.

\end{enumerate}

\begin{figure}[t!]
 \centering
 \includegraphics[scale=0.45]{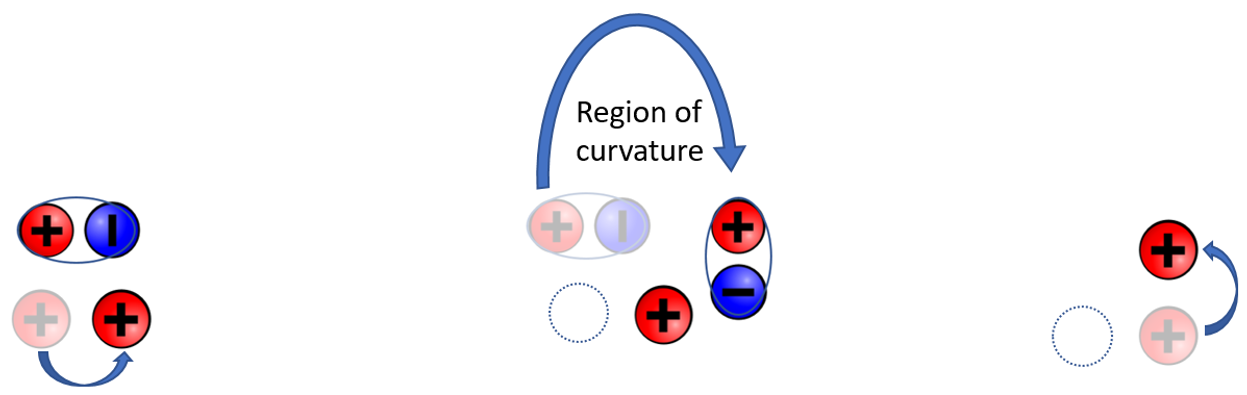}
 \caption{The mobility restriction on fractons can be lifted via virtual processes involving the parallel transport of dipoles.  First, a fracton moves in a specific direction and emits a dipole oriented in the opposite direction.  That dipole can then rotate upon being parallel transported around a closed curve.  The fracton can then reabsorb the rotated dipole, resulting in net motion of the fracton.
 }\label{fig:motion}
\end{figure}

The result of this process, as depicted in \figref{fig:motion}, is that the fracton is able to undergo net motion without the need for any other charges in the final configuration.  As such, the location of a fracton no longer defines a sharp superselection sector, as was the case in flat space, and the system can generically have nonzero hopping matrix elements moving a fracton from one site to the next, mediated by the emission and reabsorption of virtual dipoles.

For a classical system, moving a fracton may require an energy input to create dipoles, resulting in a significant potential barrier between configurations with fractons at different sites.  However, the final configuration will be energetically equivalent to the original configuration, in contrast to the behavior of fractons in flat space.  While this potential barrier can significantly inhibit the motion of fractons, the resulting behavior will not be qualitatively different from a system of ordinary particles moving in a classical potential, and the notion of a fracton is no longer sharply defined.  As we will see later, this goes hand-in-hand with a loss of the higher rank gauge invariance present in the flat-space theory.

Similar considerations to the above can be applied to several other symmetric tensor gauge theories.  For example, the vector charge theory features conservation of both vector charge and angular moment of charge, a vector quantity (in three dimensions) which we label as $\vec{L}$.  A one-dimensional particle can move in its forbidden direction at the expense of emitting a particle carrying nonzero $\vec{L}$, referred to as an $L$-particle, in the language of \refcite{deconfined}.  Upon being parallel transported around a closed curve, the $L$-particle can rotate its orientation, be reabsorbed by the vector charge, and thereby effect motion of the one-dimensional particle in its forbidden direction.

\begin{figure}[t!]
 \centering
 \includegraphics[scale=0.35]{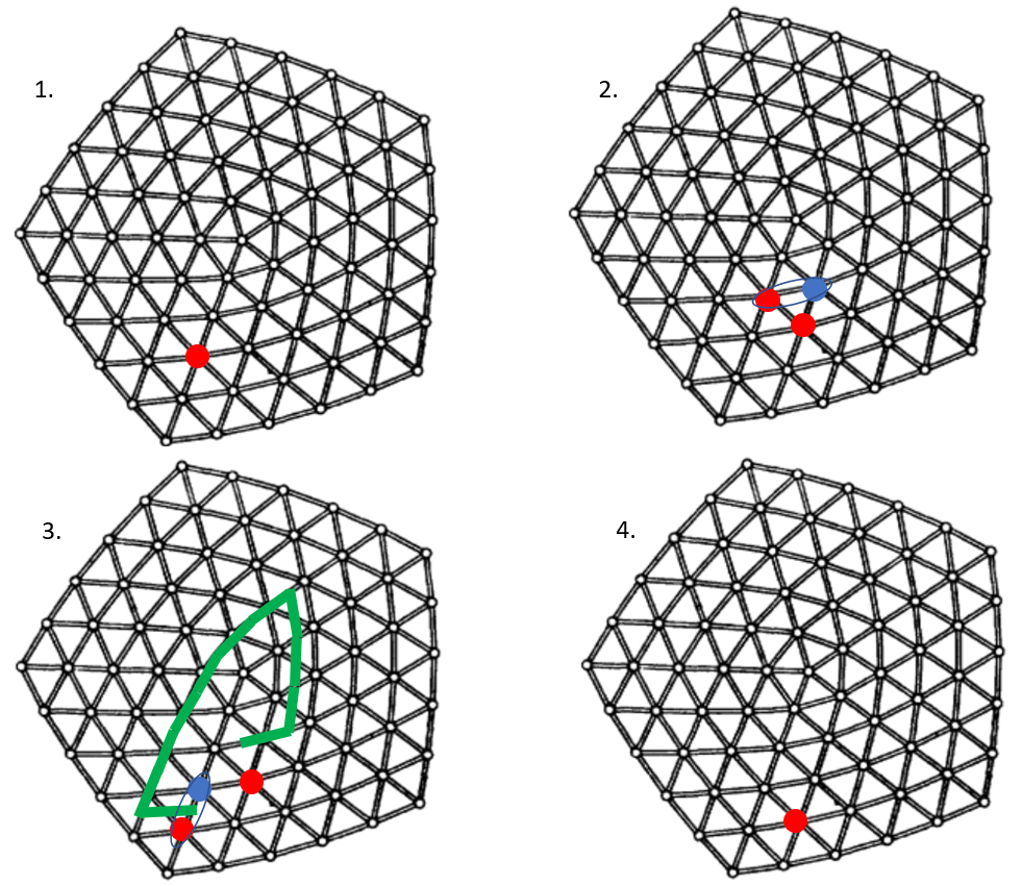}
 \caption{On a lattice, the mobility restriction on fractons is violated by a (two-dimensional) virtual dipole propagating around a disclination defect, $i.e.$ a quantized unit of curvature (and torsion), which rotates the dipole and results in net motion of the fracton.}
 \label{fig:disc}
\end{figure}

We have now seen how higher moment charge conservation laws and their associated mobility restrictions can break down in the presence of curvature.  However, in order to make connection with real physical systems, it is important to establish just how severely these restrictions are violated.  After all, most realistic systems will have some nonzero amount of curvature.  For a lattice system, this curvature can arise microscopically via dislocation and disclination defects, which are generically present in most solid samples.  A single disclination defect can lead to the breakdown of strict immobility of fractons, as depicted in \figref{fig:disc}.  However, it would be rather odd if the properties of the system could be drastically altered by the introduction of a single lattice defect.

More plausibly, we expect that the introduction of weak curvature to a system described by a symmetric tensor gauge theory should result in only mild changes to its physical properties.  In order to determine the extent to which weak curvature violates mobility restrictions in tensor gauge theories, it is simplest to study quantized curvature in the form of disclination defects, which can later be coarse-grained to make contact with a continuum description of curvature.  We work in two spatial dimensions for simplicity, but the general features will be applicable to any dimension.  The curvature-induced hopping matrix elements for a fracton are due to virtual processes such as the one depicted in \figref{fig:disc}, involving a dipole propagating around a disclination defect.  The matrix element for such a process will contain a factor of the propagator of the dipole going around the defect, which we assume is a distance $r$ away from the fracton.  If we also assume that the dipoles are gapped (i.e. carry a finite energy cost), as is generically the case in fracton models, then the hopping matrix element for a fracton will behave as:
\begin{equation}
t(r) \propto e^{- r/\xi}
\end{equation}
where $\xi$ is an effective dipole correlation length set by the energy cost $E_0$ for creating a dipole.  For a small density of disclinations ($i.e.$ small curvature), the mobility of a fracton will be dominated by dipole propagation around the nearest disclination.  Taking the disclination density to be $n_d$, the typical distance from the fracton to the nearest disclination is $r\propto n_d^{-1/2}$.  (This scaling remains the same in higher dimensions, where disclinations become higher-dimensional objects, such as the disclination lines observed in three-dimensional crystals.)  Within a coarse-grained description, the disclination density is directly proportional to the Ricci scalar: $R \propto n_d$.  Combining these pieces of information, the hopping matrix elements for a fracton behave as
\begin{equation}
t(R)\propto e^{-1/\xi \sqrt{R}}, \label{eq:hop}
\end{equation}
which results in exponentially suppressed mobility as $R\rightarrow 0$, $i.e.$ as the flat space limit is approached.  Alternatively, this can be phrased in terms of the inertial mass of a fracton\footnote{Note that, in models without Lorentz invariance (e.g. fracton models), the energy gap $E_0$ to create a particle is (in general) different from the particle's inertial mass $m$. That is, the energy of a particle with small momentum is $E(p) = E_0 + \frac{p^2}{2m} + O(p^4)$ with $E_0 \neq m$ in general.  (We use units where $c=1$.  The quantity $c^2$ tunes the ratio of the coefficients of the $B^2$ and $E^2$ terms in the Hamiltonian, which tunes the constant of proportionality for the gauge mode dispersion $\omega \sim c\,k^z$ where $z$ is the dynamical critical exponent. For the 3D scalar charge theories, $z=1$ and $c$ corresponds to the speed of light in the tensor gauge theory.)}. Within a semi-classical approximation, the effective inertial mass of a fracton can be written as
\begin{equation}
m(R) \propto t^{-1}(R)\propto e^{1/\xi\sqrt{R}}.
\end{equation}
We see that the inertial mass grows rapidly as the curvature decreases, eventually diverging in the flat space limit, recovering fracton behavior.  To have any significant motion, the curvature must satisfy $R\sim n_d\sim \xi^{-2}$.  For a large dipole gap, the effective dipole correlation length $\xi$ will be on the order of the lattice scale $a$, so the condition for significant mobility becomes $n_d\sim a^{-2}$.  At this point, the disclination density becomes comparable to the density of atoms in the solid, such that the crystalline structure is unrecognizably distorted.  For realistic solids, the effective mass will be so large that any motion of fractons will be negligible, allowing the fracton phenomenon to survive for all practical purposes in solid state systems.

The same logic as above can be carried over to other particles of restricted mobility, such as one-dimensional particles.  Motion of a one-dimensional particle in its forbidden directions requires an $L$-particle to propagate around a disclination.  Since the $L$-particles are generically gapped, the matrix element for this process will decay exponentially in the distance to the nearest disclination.  The result is that any significant motion of a one-dimensional particle in its forbidden directions requires $n_d\sim a^{-2}$, while a solid with a realistic defect density will feature negligible violations of its mobility restrictions.  For a weakly curved sample, a one-dimensional particle will still mostly move along the analogue of ``straight lines,'' which in curved space corresponds to geodesics, as we demonstrate explicitly in \appref{app:1d geodesics}.

%%%%%%%%%%%%%%%%%%%%%%%%%%%%%%%%%%%%%%%%

\section{Robustness of Fractons on Einstein Manifolds}
\label{sec:fractons on einstein}

In this section, we will argue that the traceless scalar gauge theory has well-defined subdimensional particles and maintains gauge invariance on Einstein manifolds.  However, the traceful scalar and vector charge theories do not have well-defined subdimensional particles on Einstein manifolds.

\subsection{Traceless Scalar Charge Theory}
\label{sec:traceless scalar gauge}

In this subsection, we will argue that Einstein manifolds are necessary in order for the traceless scalar charge theory to have well-defined fractons and 2D particles.  In this work, an Einstein manifold is defined to be a torsion-free manifold for which the Ricci tensor is proportional to the metric,
\begin{equation}
  R_{ab} = \tensor{R}{^c_a_c_b} = \frac{R}{D} g_{ab}. \label{eq:einstein}
\end{equation}
$\tensor{R}{^a_b_c_d}$ is the Riemann curvature tensor of the spatial manifold;
  $R = g^{ab} R_{ab}$ is the curvature scalar;
  and $D$ is the number of spatial dimensions.

As discussed in \secref{sec:curvature},
  a fracton can move only by emitting a fracton dipole with dipole moment proportional to the displacement vector of the motion.
In the traceless scalar charge theory (\secref{sec:traceless scalar}), a fracton dipole is a 2D particle
  that can only move orthogonal to its dipole moment.
When a fracton dipole traverses curved space,
  its dipole moment is parallel transported.
If a fracton dipole can traverse a small closed loop, such that its dipole moment is modified upon traversing the loop, then the fracton dipole is no longer a 2D particle since it can now move in a different direction due to its modified dipole moment.
Furthermore, the presence of such fracton dipoles in the spectrum allows a fracton to move by first emitting and then reabsorbing a dipole which changes its dipole moment, similar to the example in \figref{fig:disc}.

In curved space, when a fracton dipole traverses a small loop
  ($\vec{x} \to \vec{x}+\vec{A} \to \vec{x}+\vec{A}+\vec{B} \to \vec{x}+\vec{B} \to \vec{x}$),
  its dipole moment $P^a$ is parallel transported (i.e. rotated) according to $\delta P^a = \tensor{R}{^a_b_c_d}$~\cite{Carroll,Nakahara}.
Given the argument in the preceding paragraph,
  the existence of subdimensional particles therefore requires that the dipole moment does not rotate:
\begin{align}
\begin{split}
  & \delta P^a = \tensor{R}{^a_b_c_d} \, P^b A^c B^d = 0, \label{eq:dP} \\
\text{where } & P^a A_a = P^b B_b = 0.
\end{split}
\end{align}
We only need $\delta P^a = 0$ when $P^a A_a = P^b B_b = 0$ because fracton dipoles can only move orthogonal to their dipole moment in the traceless scalar charge theory.
We will now show that \eqnref{eq:dP} results if and only if the manifold is an Einstein manifold [\eqnref{eq:einstein}].

The Weyl tensor~\cite{Weinberg,Nakahara}\footnote{\refcite{Nakahara} has a sign mistake in front of the $\frac{1}{D-2}$.}
\begin{equation}
  C_{abcd} = R_{abcd} - \frac{1}{D-2} \Big\{\big[R_{ac}g_{bd} - (a \leftrightarrow b)\big] - (c \leftrightarrow d)\Big\} + \frac{R}{(D-2)(D-1)} (g_{ac}g_{bd} - g_{ad}g_{bc}) \label{eq:Weyl}
\end{equation}
vanishes identically in $D=3$ spatial dimensions.
This implies that Einstein manifolds
  have a Riemann curvature tensor given by
\begin{equation}
  R_{abcd} = \frac{R}{D(D - 1)} (g_{ac} g_{bd} - g_{ad} g_{bc}) \label{eq:R}
\end{equation}
for spatial dimensions $D \leq 3$.
Plugging the above equation into \eqnref{eq:dP} gives
\begin{align}
\begin{split}
  \delta P^a &= \frac{R}{6} (g_{bd} P^b A^a B^d - g_{bc} P^b A^c B^a) \\
    &= 0,
\end{split}
\end{align}
since $P^a A_a = P^b B_b = 0$.
Thus, Einstein manifolds in three dimensions obey \eqnref{eq:dP}.

To show that all 3D manifolds obeying \eqnref{eq:dP} are Einstein manifolds,
  first suppose a manifold obeys \eqnref{eq:dP}.
Next, decompose the metric into frame fields $e_i^a(x)$ (with $i=1,2,3$).
(In the rest of this subsection, we will reserve the letters $i$, $j$, and $k$ to index the frame fields.)
Frame fields are a set of orthonormal vectors at each point in space
  ($e_i^a g_{ab} e_j^b = \delta_{ij}$)
  with inverses $e_a^i$ (where $e^i_a e_i^b = \delta_a^b$).
The metric can then be decomposed into frame fields as $g_{ab} = e^i_a \delta_{ij} e^j_b$.
Now consider a component of the Ricci tensor in the frame field basis:
\begin{align}
  R_{bd} \, e_1^b e_2^d
    &= \tensor{R}{^a_b_a_d} \, e_1^b e_2^d \nonumber\\
    &= R_{abcd} \underbrace{e_i^a \delta^{ij} e_j^c}_{g^{ac}} e_1^b e_2^d \nonumber\\
    &= R_{abcd} \, \delta^{ij} e_i^a e_1^b e_j^c e_2^d \nonumber\\
    &= R_{abcd} \, e_1^a e_1^b e_1^c e_2^d
     + R_{abcd} \, e_2^a e_1^b e_2^c e_2^d 
     + R_{abcd} \, e_3^a e_1^b e_3^c e_2^d \label{eq:Ricci 123}\\
    &= 0. \nonumber
\end{align}
To obtain \eqnref{eq:Ricci 123}, we have explicitly summed over $i,j=1,2,3$.
The first two terms in \eqnref{eq:Ricci 123} are zero due to antisymmetries of the Riemann curvature tensor ($R_{abcd} = - R_{bacd} = - R_{abdc}$);
  the third term is zero due to \eqnref{eq:dP}
  and the orthogonality of the frame fields $e^a_i$.
However, note that throughout this calculation, the choice of orthogonal frames was arbitrary.
Thus, $R_{ab} A^a B^b = 0$ for any pair of vectors $A^a$ and $B^b$ that are orthogonal ($i.e.$ $g_{ab} A^a B^b = 0$).
This implies the Einstein condition [\eqnref{eq:einstein}]
  (see \appref{app:proveEinstein} for a proof).

We will now show that in addition to having well-defined subdimensional particles, the traceless scalar charge theory (\secref{sec:traceless scalar}) is in fact also gauge invariant on Einstein manifolds.
First, we must generalize the flat-space gauge transformations [\eqsref{eq:scalar gauge} and \eqref{eq:scalar traceless gauge}] to curved space.
This can be done by simply adding metrics ($\delta_{ij} \to g_{ij}$) and covariant derivatives ($\partial_i \to \nabla_i$).
However, it can also be systematically derived from the curved-space generalization of the Gauss law and traceless constraints ($\nabla_a \nabla_b E^{ab} = g_{ab} E^{ab} = 0$) using Poisson brackets as follows:
\begin{align}
\begin{split}
  A_{ij}(t,\vx) &\stackrel{\lambda}{\to} A_{ij}(t,\vx) - \int_{\vx'}
    \big\{ A_{ij}(t,\vx), \underbrace{\nabla'_a \nabla'_b E^{ab}(t,\vx')}_{\rho(t,\vx')} \big\} \lambda(t,\vx') \\
    &= A_{ij}(t,\vx) + \nabla_i \nabla_j \lambda(t,\vx),
\end{split} \label{eq:scalar gauge lambda}\\
\begin{split}
  A_{ij}(t,\vx) &\stackrel{\mu}{\to} A_{ij}(t,\vx) - \int_{\vx'}
    \big\{ A_{ij}(t,\vx), \underbrace{g_{ab} E^{ab}(t,\vx')}_{\text{traceless}} \big\} \mu(t,\vx') \\
    &= A_{ij}(t,\vx) + g_{ij} \mu(t,\vx),
\end{split} \label{eq:scalar gauge mu}
\end{align}
where the Poisson bracket is
\begin{align}
  \{ A_{ij}(t,\vx), E^{ab}(t,\vx') \} = - \tfrac{1}{2} (\delta_i^a \delta_j^b + \delta_i^b \delta_j^a) \delta^3(\vx - \vx').
\end{align}

The flat-space magnetic field [\eqnref{eq:scalar traceless mag}] extends naturally to curved-space:
\begin{equation}
  B^{ij} = \frac{1}{2} g^{jc} \epsilon^{iab} \nabla_a A_{bc} + (i \leftrightarrow j),
\end{equation}
where $\epsilon^{ijk}$ is the Levi-Civita tensor.
We can now show that the magnetic field is invariant under the gauge transformation [\eqnref{eq:scalar gauge lambda}]:
\begin{align}
  B^{ij} &\stackrel{\lambda}{\to} \frac{1}{2} B^{ij} + \frac{1}{2} g^{jc} \epsilon^{iab} \nabla_a \nabla_b \nabla_c \lambda + (i \leftrightarrow j) \nonumber\\
  &= \frac{1}{2} B^{ij} + \frac{1}{4} g^{jc} \epsilon^{iab} [\nabla_a, \nabla_b] \nabla_c \lambda + (i \leftrightarrow j) \nonumber\\
  &= \frac{1}{2} B^{ij} - \frac{1}{4} g^{jc} \epsilon^{iab} \tensor{R}{^k_c_a_b} \nabla_k \lambda + (i \leftrightarrow j) \label{eq:B1}\\
  &= \frac{1}{2} B^{ij} - \frac{R}{24} g^{jc} \epsilon^{iab} (\delta_a^k g_{cb} - \delta_b^k g_{ca}) \nabla_k \lambda + (i \leftrightarrow j) \label{eq:B2}\\
  &= \frac{1}{2} B^{ij} - \frac{R}{24} (\epsilon^{iaj} \nabla_a \lambda - \epsilon^{ijb} \nabla_b \lambda) + (i \leftrightarrow j) \nonumber\\
  &= B^{ij}. \label{eq:B3}
\end{align}
\eqnref{eq:B1} results from the torsion-free\footnote{A non-zero torsion tensor $\tensor{T}{^k_a_b} = \Gamma^k_{ab} - \Gamma^k_{ba}$ would result in a $-\frac{1}{4} g^{jc} \epsilon^{iab} \tensor{T}{^k_a_b} \nabla_k \nabla_c \lambda + (i \leftrightarrow j)$ term in \eqnref{eq:B1}, which would break gauge invariance.}
  curvature tensor identity $[\nabla_a,\nabla_b]V_c = -\tensor{R}{^k_c_a_b} V_k$;
  \eqnref{eq:B2} follows from \eqnref{eq:R},
  which resulted from the assumption of an Einstein manifold;
  and \eqnref{eq:B3} results from the $i \leftrightarrow j$ symmetrization.

The magnetic field is also invariant under the local symmetry [\eqnref{eq:scalar gauge mu}] necessitated by the traceless condition,
\begin{align}
  B^{ij} &\stackrel{\mu}{\to} \frac{1}{2} B^{ij} + \frac{1}{2} g^{jc} \epsilon^{iab} g_{bc} \nabla_a \mu + (i \leftrightarrow j) \label{eq:B4}\\
  &= \frac{1}{2} B^{ij} + \frac{1}{2} \epsilon^{iaj} \nabla_a \mu + (i \leftrightarrow j) \nonumber\\
  &= B^{ij}. \label{eq:B5}
\end{align}
Here, \eqnref{eq:B4} made use of metric compatibility $\nabla_a g_{bc} = 0$,
  while \eqnref{eq:B5} results from the $i \leftrightarrow j$ symmetrization.

\subsection{Traceful Scalar and Vector Charge Theories}
\label{sec:traceful}

In \secref{sec:traceless scalar gauge}, we argued that the traceless scalar charge theory has well-defined fractons on Einstein manifolds.  In the argument, it was critical that the dipole charge was a 2D particle (so that we could make use of the constraint in \eqnref{eq:dP}).  In the traceful scalar and vector charge theories, the dipole charge is a fully-mobile particle.  Thus, the dipole can traverse arbitrary loops, which will parallel-transport its dipole moment, and result in full mobility for the scalar charges (fractons in flat space) in the scalar charge theory (as explained in \secref{sec:curvature}) and vector charges (1D particles in flat space) in the vector charge theory.  This argument applies in any number of spatial dimensions\footnote{\textit{A priori}, the two-dimensional vector charge theory could have been an exception to this logic, since the conserved quantity is a scalar.}, and this issue occurs even for the most symmetric curved manifolds: the sphere and hyperbolic space.  We therefore do not expect these theories to be gauge invariant in the presence of curvature.

For example, the gauge invariance of the magnetic field in the traceless scalar theory relied on the fact that the magnetic tensor is symmetric ($B^{ij} = B^{ji}$), which is not the case for the traceful scalar theory.  And in \appref{app:traceful 2d vector}, we explicitly show that the 2D vector charge theory is not gauge invariant on manifolds with constant curvature. We have also checked that the 3D traceful vector charge theory is not gauge invariant even on Einstein manifolds with constant curvature, with details of this calculation provided in the accompanying Mathematica notebook~\cite{notebook}.

\subsection{Traceless Vector Charge Theory}
\label{sec:tracelessVector}

In the traceless vector charge theory (reviewed in \secref{tracelessvectorreview}), the vector charges are fractons, while the $L$-particle bound states are one-dimensional~\cite{genem}.  It therefore seems possible that this theory may evade some of the difficulties encountered in theories with two-dimensional or fully mobile dipoles.  Nevertheless, we find that gauge invariance and mobility restrictions are only maintained for a certain special class of manifolds.  Specifically, if the magnetic field tensor [generalized from \eqnref{eq:tracelessVecB}] is defined as
\begin{equation}
  B^{ij} = \frac{1}{2} \epsilon^{iab} g^{je} g^{cd} \nabla_c (\nabla_e \nabla_a A_{bd} - \nabla_a \nabla_d A_{be}) + (i \leftrightarrow j) \label{eq:traceless vector B}
\end{equation}
then the magnetic field is gauge invariant on Einstein manifolds with a constant curvature scalar ($\nabla_i R = 0$). The gauge transformation is $A_{ij} \to A_{ij} + \frac{1}{2} (\nabla_i \lambda_j + \nabla_j \lambda_i)$, which is generalized from \eqnref{eq:vectorGauge}. If the constant curvature scalar is not constant, then $B^{ij}$ is not gauge invariant (in general).  We do not have a crisp physical explanation for why constant curvature is necessary for gauge invariance.

We obtained this result by using a Mathematica notebook~\cite{notebook} to evaluate $B^{ij}$ under the gauge transformation.
We have not checked the result by hand due to the very lengthy mathematical expressions that are involved.

One may wonder if other orderings of the covariant derivatives in $B^{ij}$ could result in a magnetic field tensor that is gauge invariant on Einstein manifolds without a constant curvature scalar ($\nabla_i R \neq 0$).
However, in our Mathematica notebook~\cite{notebook},
  we have checked that no linear combination of different covariant derivative orderings can result in a gauge invariant magnetic field.
The various linear combinations result in a dimension-$12$ vector space of possible magnetic field tensors.
We then evaluate $B^{ij}$ with $A_{ij} = \frac{1}{2} (\nabla_i \lambda_j + \nabla_j \lambda_i)$.
Next, we reorder all covariant derivatives (via the identity $[\nabla_a,\nabla_b]V_c = -\tensor{R}{^k_c_a_b} V_k$)
  into a canonical order:
  e.g. $\nabla_2 \nabla_1 \lambda_3 \to [\nabla_2, \nabla_1] \lambda_3 + \nabla_1 \nabla_2 \lambda_3$.
After choosing a coordinate system where $g^{ij} = \delta^{ij}$ at a given point,
  we simplify the resulting expression to a sum of terms,
  such as
  $\beta_{1,3,2} \nabla_1 R \, \nabla_2 \lambda_3$ or
  $\alpha_{3,1,2} \nabla_1 \nabla_2 \nabla_2 \nabla_3 \lambda_2$,
  where $\alpha$ and $\beta$ are coefficients for the different covariant derivative orderings.
We then assume that all expressions of the form 
  $S_{abc} = \nabla_a \nabla_b R \, \lambda_c$ (with $a \leq b$) or
  $T_{abc} = \nabla_a R \, \nabla_b \lambda_c$ or $U_{abc} = R \, \nabla_a \nabla_b \lambda_c$ (with $a \leq b$) or
  $V_{abcde} = \nabla_a \nabla_b \nabla_c \nabla_d \lambda_e$ (with $a \leq b \leq c \leq d$)
  are independent.
Thus, in order for $B^{ij}$ to be gauge invariant,
  we must choose the twelve $\alpha$ and $\beta$ coefficients such that all of the $S_{abc}$, $T_{abc}$, $U_{abc}$, and $V_{abcde}$ expressions cancel.
This just requires solving a linear system of equations.
However, the only solution is $B^{ij} = 0$.
Therefore, no linear combination of the different orderings of the covariant derivatives in $B^{ij}$ results in a gauge invariant magnetic field.

% The gauge transformations are
% \begin{align}
%   A_{ij} &\stackrel{\lambda}{\to} A_{ij} + \frac{1}{2} (\nabla_i \lambda_j + \nabla_j \lambda_i) \\
%   A_{ij} &\stackrel{\mu}{\to} A_{ij} + g_{ij} \mu
% \end{align}
% The first is the curved space generalization of the flat-space gauge transformation for a vector charge theory: \eqnref{eq:vectorGauge}.
% The second is the symmetry that results from the traceless condition and also appeared in the traceless scalar theory [\eqnref{eq:scalar gauge mu}].

%%%%%%%%%%%%%%%%%%%%%%%%%%%%%%%%%%%%%%%%

\section{Conclusions}
\label{cncls}

\begin{table}
\center
\begin{tabular}{r@{\hspace{2pt}}l@{\hspace{2pt}}l@{\hspace{2pt}}l|c|c}
  \multicolumn{4}{c|}{symmetric tensor gauge theory} & gauge invariant manifold & reference \\ \hline
  3D & gapless & traceless & scalar & Einstein & \secref{sec:traceless scalar gauge} \\
  3D & gapless & traceless & vector & Einstein with constant curvature & \secref{sec:tracelessVector} \\
  2D & gapped & traceless & scalar & constant curvature & Eq. (5) of \refcite{gromov} \\
  2D & gapless & traceless & scalar & constant curvature & $\dagger$ \\
  any-D & gapless & traceful & scalar & flat & \secref{sec:traceful} \\
  any-D & gapless & traceful & vector & flat & \secref{sec:traceful}, \appref{app:traceful 2d vector} for 2D
\end{tabular}
\caption{A summary of the kinds of manifold for which each symmetric tensor gauge theory maintains gauge invariance and sharp mobility restrictions.  Einstein manifolds obey \eqnref{eq:einstein} while constant curvature manifolds have a constant curvature scalar $\partial_i R = 0$; flat manifolds have no curvature ($\tensor{R}{^a_b_c_d} = 0$); all of these manifolds are torsion-free.
$\dagger$ The gapped and gapless 2D traceless scalar charge theories have the same magnetic field and are gauge invariant under the same class of manifolds.}
\label{tab:summary}
\end{table}

In this work, we have argued that arbitrary curvature can lead to violations of the mobility restrictions associated with subdimensional particles (\secref{sec:curvature}).  Curvature can grant fractons (and other subdimensional particles) full mobility and often results in a loss of gauge invariance.  However, if the curvature is weak, the fracton hopping strength is exponentially small [\eqnref{eq:hop}].  Furthermore, in certain theories, subdimensional particles are robust on Einstein manifolds, for which the curvature can be described by a spatially-dependent cosmological constant $\Lambda(x)$ [\eqnref{eq:einstein}].  This is the case for the traceless scalar charge theory, which we showed is gauge invariant on Einstein manifolds (\secref{sec:traceless scalar gauge}).  However, not all fracton theories remain gauge invariant on Einstein manifolds.  For example, we found that it was important that the dipole charge is not fully mobile (\secref{sec:traceful}), which implies that the traceful scalar and vector charge theories are not gauge invariant even on the maximally-symmetric curved manifolds (in any number of dimensions): the sphere or hyperbolic space.  See \tabref{tab:summary} for a summary of what kind of manifold each theory is gauge invariant under.

In two spatial dimensions, Eq.~(5) of~\refcite{gromov} shows that the gapped traceless scalar charge theory~\cite{prem2} is gauge invariant only if the curvature is constant (i.e. has no dependence on space or time).  Since it has the same expression for the magnetic field, the 2D {\it gapless} traceless scalar charge theory is also gauge invariant when the curvature is constant.  We do not know of a physical explanation, such as the one in \secref{sec:curvature}, for why constant curvature is necessary in these theories or the 3D traceless vector theory.

Although we only considered curved space in this work, it would be very interesting if the tensor gauge theories could be generalized to curved spacetime.  However, this is very non-trivial since although the flat-space tensor gauge theories in this work have rotation symmetry, they do not have Lorentz symmetry.
There is a physical reason for this: the existence of subdimensional particles is not Lorentz invariant.
For example, after a Lorentz boost, an immobile fracton would become a particle that is constrained to move at a fixed velocity, and such particles are not present in the theories considered in this work.

In \refcite{shirley,foliatedEntanglement,foliatedExcitations}, it was shown that many of the gapped fracton theories can be understood as having a foliation or layered structure.  For example, some properties, such as the 2D particles or the two leading terms in the entanglement entropy~\cite{foliatedEntanglement}, can be understood by approximating the fracton order as three or more decoupled stacks of 2D topological orders.  If these layers are curved, then the gapped fracton phase is described on a curved lattice with a natural set of curved surfaces.  Einstein manifolds appear to be the analog of this for rotation-symmetric fracton phases.  Einstein manifolds also have a natural set of surfaces: the surfaces that the 2D particles can move along.

An interesting consequence of the weak mobility acquired by fractons/sub-dimensional particles pertains to their dynamical behaviour. The closed system quantum dynamics for gapped fracton models was studied in Ref.~\cite{prem}, where it was shown that the mobility of fractons is exponentially suppressed in the inverse temperature $T$, set by the bath of composite mobile excitations. Following the discussion in this work, for symmetric tensor gauge theories we expect that the dynamical behaviour of an isolated quantum system will be dictated by both the temperature $T$ and the curvature $R$, since the latter weakly endows fractons with mobility. For traceless scalar charge theories defined on Einstein manifolds, the analysis of Ref.~\cite{prem} will carry over exactly since curvature effects do not change any of the restrictions on the gapped excitations for this class of systems. More generally, the mobility of sub-dimensional excitations can grow with the curvature; equivalently, the time-scale for thermalization will decrease with increasing curvature while still being suppressed by the inverse temperature. For weak enough curvature though, the glassy dynamics seen in gapped fracton models~\cite{chamon,prem} should persist even for the class of symmetric tensor gauge theories discussed here. 

Recently, a gauge theoretic construction for phases where excitations are created at the ends of fractal operators, such as those present in Haah's code~\cite{haah}, was discussed~\cite{haahU1,BulmashFractal}. In this class of U(1) theories, there are {\it no} topologically non-trivial excitations which are mobile; i.e. all excitations carrying a topological charge are strictly immobile. Whether this class of models can retain its characteristic immobility in the presence of curvature remains to be seen.

\section*{Acknowledgments}

We acknowledge stimulating conversations and correspondence with Brian Swingle, Mike Hermele, Will Jay, Rahul Nandkishore, Danny Bulmash, Andrey Gromov, and John McGreevy.
KS is grateful for support from the NSERC of Canada, the Center for Quantum Materials at the University of Toronto, and the Walter Burke Institute for Theoretical Physics at Caltech.  MP is supported partially by a Simons Investigator Award to Leo Radzihovsky and partially by the NSF Grant 1734006. This material is based in part (AP) upon work supported by the Air Force Office of Scientific Research under award number FA9550-17-1-0183. 

%---------------------------------------
%---------------------------------------

\appendix
\renewcommand*{\thesection}{\Alph{section}} % fix appendix text in the table of contents: https://tex.stackexchange.com/a/26348

\section{Einstein Manifold Proof}
\label{app:proveEinstein}

Suppose that for any pair of orthogonal vectors $A^a$ and $B^b$,
  we have $R_{ab} A^a B^b=0$ where $R_{ab}$ is a symmetric tensor:
\begin{equation}
  g_{ab} A^a B^b = 0 \implies R_{ab} A^a B^b=0 \label{eq:assumption}
\end{equation}
We will now prove that the above assumption implies that $R_{ab} \propto g_{ab}$.

We will begin by diagonalizing $g_{ab}$ by a set of invertible frame fields $e^a_i$:
\begin{equation}
  e^a_i g_{ab} e^b_j = \delta_{ij} \label{eq:basis}
\end{equation}
In this appendix, we reserve the indices $i$ and $j$ for the Euclidean frame field coordinates.

Now note that the linear combinations of the following set of tensors span all symmetric tensors ($T^{ab}$ where $T^{ab} = T^{ba}$):
\begin{align}
  g^{ab} & \nonumber\\
  Q^{ab}_{ij} &= e^a_i e^b_j + e^a_j e^b_i \nonumber\\
  S^{ab}_{ij} &= e^a_i e^b_i - e^a_j e^b_j \nonumber\\
    &= \frac{1}{2} (e^a_i + e^a_j)(e^b_i - e^b_j) 
     + \frac{1}{2} (e^a_i - e^a_j)(e^b_i + e^b_j) \label{eq:Sab}
\end{align}
where $i \neq j$.
One should think of $(e^a_1, e^a_2, .., e^a_D)$ as a basis of vectors.
Then $Q^{ab}_{ij}$ spans all off-diagonal components of symmetric tensors,
  while $S^{ab}_{ij}$ and $g^{ab}$ together span all diagonal components.

Now note that for any $i \neq j$,
  $g_{ab} A^a B^b = 0$ when
  $A^a = e_i^a$ and $B^b = e_j^b$ or when
  $A^a = e^a_i + e^a_j$ and $B^b = e^b_i - e^b_j$.
Our initial assumption [\eqnref{eq:assumption}]
  then implies that
  $R_{ab} A^a B^b = 0$ for these pairs of $A^a$ and $B^b$,
  which means that $R_{ab} Q^{ab}_{ij} = R_{ab} S^{ab}_{ij} = 0$ for all $i \neq j$ (by expressing $S^{ab}_{ij}$ as in \eqnref{eq:Sab}).
Therefore, $R_{ab}$ is orthogonal to $Q^{ab}_{ij}$ and $S^{ab}_{ij}$.
This implies that $R_{ab} \propto g_{ab}$
  since $Q^{ab}_{ij}$, $S^{ab}_{ij}$, and $g^{ab}$ span all symmetric tensors.
This completes the proof.

\section{Motion of one-dimensional particles along geodesics}
\label{app:1d geodesics}

Here, we demonstrate that the one-dimensional particles in the vector charge theory are confined to move along geodesics, to the extent that their mobility restrictions are obeyed.

The vector charges of this theory, which in flat space can only move in the direction of their charge vector, satisfy the generalized continuity equation~\cite{genem}
\beq
\label{continuity}
\partial_0 \rho^j + \partial_j J^{ij} = 0\quad \forall \, i,
\eeq
where $J^{ij} = J^{ji}$ is the symmetric current tensor.

For the class of metrics considered in this paper $i.e.$ those with only spatial curvature, the Christoffel symbols
\beq
\Gamma_{\mu \nu}^\sigma = \frac{1}{2} g^{\sigma \rho} \left(\partial_\mu g_{\nu \rho} + \partial_\nu g_{\mu \rho} - \partial_\rho g_{\mu \nu} \right)
\eeq
simplify so that
\beq
\Gamma^0_{\mu \nu} = \Gamma^\sigma_{\mu 0} = \Gamma^\sigma_{0\nu} = 0,  
\eeq
and only the spatial components $\Gamma^{k}_{ij}$ can be non-vanishing.
(Recall that we use Greek letters for spacetime indices and latin letters for spatial indices.)

The generalization of the continuity equation~\eqref{continuity} on curved manifolds is straightforward:
\beq
\nabla_0 \rho^i + \nabla_j J^{ij} = 0\quad \forall \, i,
\eeq
and can be re-written in terms of the connection co-efficients as
\beq
\label{gencont}
\partial_0 \rho^i + \partial_j J^{ij} + \Gamma^i_{jk} J^{kj} + \Gamma^j_{jk} J^{ik} = 0 \quad \forall \, i.
\eeq

Let us now consider the motion of a point-particle along a path $\mathbf{s}(t)$, for which the charge density is ($\forall\,i$)
\beq
\rho^i(t,\vx) = r^i(t)\,\delta^3(\vx- \mathbf{s}(t)),
\eeq
where $r^i(t)$ is the charge vector.
Similarly, the current density takes the form
\beq
J^{ij}(t,\vx) = \gamma^{ij}(t)\,\delta^3(\vx - \mathbf{s}(t)),
\eeq
where $\gamma^{ij} = \gamma^{ji}$ is also symmetric. Eq.~\eqref{gencont} thus becomes ($\forall\,i$)
\beq
\big[\partial_0 r^i(t) + \Gamma^i_{jk} \gamma^{kj}(t) + \Gamma^j_{jk} \gamma^{ik}(t) \big]\,\delta^3(\vx - \mathbf{s}(t)) + \big[r^i(t)\,\partial_0 s^j(t) + \gamma^{ij}(t)  \big]\,\partial_j \delta^3(\vx - \mathbf{s}(t)) = 0,
\eeq
which is satisfied if
\begin{align}
\partial_0 r^i(t) &= -\Gamma^i_{jk}(\mathbf{s}(t))\,\gamma^{kj}(t) - \Gamma^j_{jk}(\mathbf{s}(t))\,\gamma^{ik}(t) \label{coupled1} \\
\gamma^{ij}(t) &= -r^i(t)\,\partial_0 s^j(t) \label{coupled2}.
\end{align}
Since $\gamma^{ij}$ is symmetric, Eq.~\eqref{coupled2} implies that 
\beq
\partial_0 s^j(t) = \alpha^{-1} r^i(t) \label{eq:sr}
\eeq
for some arbitrary constant $\alpha$.
Therefore,
\beq
\gamma^{ij}(t) = -\frac{1}{\alpha} r^i(t)\,r^j(t)
\eeq
Plugging this form back into Eq.~\eqref{coupled1}, we find that
\beq
\label{pseudogeodesic}
\partial_0 r^i(t) = \frac{1}{\alpha} \Gamma^i_{jk}(\mathbf{s}(t))\,r^j(t)\,r^k(t) + \frac{1}{\alpha} \Gamma^j_{jk}(\mathbf{s}(t))\,r^k(t)\,r^i(t).
\eeq
This is immediately reminiscent of the geodesic equation; in fact, if not for the second term on the right, it is precisely the geodesic equation. To bring Eq.~\eqref{pseudogeodesic} into a more familiar form, we rescale $r^i(t)$:
\begin{align}
r^i(t) &= u^i(t) \int dt\, \Gamma^j_{jk}(\mathbf{s}(t))\, \frac{r^k(t)}{\alpha} \nonumber \\
&= u^i(t) \int dt\, \Gamma^j_{jk}(\mathbf{s}(t))\, \partial_0 s^k(t) \label{eq:renorm1} \\
&= u^i(t) \int dt\, \partial_k \log \left(\sqrt{|\tilde{g}(\vx)|} \right)_{\vx = \mathbf{s}(t)}\,\partial_0 s^k(t) \label{eq:renorm2} \\
&= u^i(t) \log\left(\sqrt{c\,|\tilde{g}(\mathbf{s}(t))|} \right),
\end{align}
where $c$ is some constant and $|\tilde{g}_{ij}|$ is the determinant of the time-invariant spatial part of the metric. \eqnref{eq:renorm1} follows from \eqnref{eq:sr}, and \eqnref{eq:renorm2} follows from the identity $\Gamma^\mu_{\mu \nu} = \partial_\nu \log\left(|g|\right)$.

Eq.~\eqref{pseudogeodesic} then becomes
\beq
\partial_0 u^i(t) = \frac{1}{\alpha} \log\left(\sqrt{c\, |\tilde{g}(\mathbf{s}(t))|} \right) \Gamma^i_{jk}(\mathbf{s}(t))\,u^j(t)\,u^k(t).
\eeq
By further reparametrizing $t$ such that $u^i(t) = \tilde{u}^i(\tau(t))$ where
\beq
\tau(t) = - \alpha \int dt\, \frac{1}{\alpha} \log\left(\sqrt{c\, \tilde{g}(\mathbf{s}(t))} \right),
\eeq
we find that the particles obey the geodesic equation:
\beq
\partial_\tau \tilde{u}^i(\tau) = -\Gamma^i_{jk}(\mathbf{s}(\tau))\,\tilde{u}^j(\tau)\,\tilde{u}^k(\tau).
\eeq

Hence, we have demonstrated that the only point-particle solution to the generalized charge conservation equation~\eqref{gencont} on a curved manifold is 
\begin{align}
    \rho^i(t) &= \alpha\,\partial_0s^i(t)\,\delta^3(\vx - \mathbf{s}(t)), \nonumber \\
    J^{ij}(t) &= -\alpha\,\partial_0s^i(t)\,\partial_0s^j(t)\,\delta^3(\vx - \mathbf{s}(t)),
\end{align}
which describes the motion of a particle confined to move along the geodesics of the curved space. This discussion pertains only to the motion of a single particle as dictated by the continuity equation; as discussed in the main text, generically these particles will be able to stray from their geodesic path by emitting mobile dipole excitations, which, when parallel transported, will change their orientation, and reabsorbing them.

\section{2D Vector Charge Theory}
\label{app:traceful 2d vector}

In this appendix, we show that the 2D vector charge theory~\cite{deconfined} is not gauge invariant on manifolds with constant nonzero curvature.

The magnetic field and gauge transformation on curved space are
\begin{align}
  B &= \epsilon^{ai} \epsilon^{bj} \nabla_a \nabla_b A_{ij} \\
  A_{ij} &\to A_{ij} + \nabla_i \lambda_j + \nabla_j \lambda_i
\end{align}
and the Hamiltonian is $H=\int \frac{1}{2}(E^2+B^2)$ as usual.
The order of $\nabla_a \nabla_b$ in $B$ does not matter since $\epsilon^{ai} \epsilon^{bj} A_{ij}$ is symmetric under $(a \leftrightarrow b)$.

Under the gauge transformation, the magnetic field transforms as follows:
\begin{align}
B &\stackrel{\lambda}{\to}
  B + \epsilon^{ai} \epsilon^{bj} \nabla_a \nabla_b (\nabla_j \lambda_i + \nabla_i \lambda_j) \nonumber\\
  &= B + \epsilon^{ai} \epsilon^{bj} [\nabla_a, \nabla_b] \nabla_i \lambda_j + \epsilon^{ai} \epsilon^{bj} \nabla_b [\nabla_a, \nabla_i] \lambda_j \nonumber\\
  &= B - \epsilon^{ai} \epsilon^{bj} \Big[ \underbrace{\frac{R}{2} (\delta_a^c g_{ib} - \delta_b^c g_{ia})}_{\tensor{R}{^c_i_a_b}} \nabla_c \lambda_j + \underbrace{\frac{R}{2}  (\delta_a^c g_{jb} - \delta_b^c g_{ja})}_{\tensor{R}{^c_j_a_b}} \nabla_i \lambda_c \Big]
   - \epsilon^{ai} \epsilon^{bj} \nabla_b \Big[ \underbrace{\frac{R}{2} (\delta_a^c g_{ji} - \delta_i^c g_{ja})}_{\tensor{R}{^c_j_a_i}} \lambda_c \Big] \label{eq:2d vec B1}\\
  &= B - \frac{R}{2} \epsilon^{ai} \epsilon^{bj} \big(g_{ib} \nabla_a \lambda_j - g_{ia} \nabla_b \lambda_j + g_{jb} \nabla_i \lambda_a - g_{ja} \nabla_i \lambda_b + g_{ji} \nabla_b \lambda_a - g_{ja} \nabla_b \lambda_i \big) \label{eq:2d vec B2}\\
%   &= B - \frac{R}{2} \epsilon^{ai} \epsilon^{bj} (g_{ba} \nabla_i \lambda_j - g_{ib} \nabla_a \lambda_j) \nonumber\\
  &= B - R \, g^{aj} \nabla_a \lambda_j
\end{align}
\eqnref{eq:2d vec B1} makes use of the fact that all 2D manifolds satisfy $R_{abcd} = \frac{R}{2} (g_{ac}g_{bd} - g_{ad} g_{bc})$ where $R=\tensor{R}{^a^b_a_b}$ is the curvature scalar.
In \eqnref{eq:2d vec B2}, we assumed that the spatial curvature is constant (i.e. $R$ has no space-time dependence).
Therefore, the magnetic field is not gauge invariant on curved spaces with constant nonzero curvature.

%---------------------------------------
%BIBLIOGRAPHY
%---------------------------------------

\newpage 

\bibliography{curvedFractons}

\begin{thebibliography}{10}
\expandafter\ifx\csname url\endcsname\relax
  \def\url#1{\texttt{#1}}\fi
\expandafter\ifx\csname urlprefix\endcsname\relax\def\urlprefix{URL }\fi
\expandafter\ifx\csname href\endcsname\relax
  \def\href#1#2{#2} \def\path#1{#1}\fi

\bibitem{Laughlin}
R.~B. Laughlin, Anomalous quantum hall effect: An incompressible quantum fluid
  with fractionally charged excitations, Phys. Rev. Lett. 50 (1983) 1395--1398.
\newblock \href {https://doi.org/10.1103/PhysRevLett.50.1395}
  {\path{doi:10.1103/PhysRevLett.50.1395}}.

\bibitem{wenniu}
X.~G. Wen, Q.~Niu, Ground-state degeneracy of the fractional quantum hall
  states in the presence of a random potential and on high-genus riemann
  surfaces, Phys. Rev. B 41 (1990) 9377--9396.
\newblock \href {https://doi.org/10.1103/PhysRevB.41.9377}
  {\path{doi:10.1103/PhysRevB.41.9377}}.

\bibitem{mooreread}
G.~Moore, N.~Read, Nonabelions in the fractional quantum hall effect, Nuclear
  Physics B 360~(2) (1991) 362 -- 396.
\newblock \href {https://doi.org/10.1016/0550-3213(91)90407-O}
  {\path{doi:10.1016/0550-3213(91)90407-O}}.

\bibitem{wen2002}
X.-G. Wen, Quantum orders and symmetric spin liquids, Phys. Rev. B 65 (2002)
  165113.
\newblock \href {https://doi.org/10.1103/PhysRevB.65.165113}
  {\path{doi:10.1103/PhysRevB.65.165113}}.

\bibitem{balents2002}
L.~Balents, M.~P.~A. Fisher, S.~M. Girvin, Fractionalization in an easy-axis
  kagome antiferromagnet, Phys. Rev. B 65 (2002) 224412.
\newblock \href {https://doi.org/10.1103/PhysRevB.65.224412}
  {\path{doi:10.1103/PhysRevB.65.224412}}.

\bibitem{moessner}
R.~Moessner, S.~L. Sondhi, E.~Fradkin, Short-ranged resonating valence bond
  physics, quantum dimer models, and ising gauge theories, Phys. Rev. B 65
  (2001) 024504.
\newblock \href {https://doi.org/10.1103/PhysRevB.65.024504}
  {\path{doi:10.1103/PhysRevB.65.024504}}.

\bibitem{levinwen}
M.~A. Levin, X.-G. Wen, String-net condensation: A physical mechanism for
  topological phases, Phys. Rev. B 71 (2005) 045110.
\newblock \href {https://doi.org/10.1103/PhysRevB.71.045110}
  {\path{doi:10.1103/PhysRevB.71.045110}}.

\bibitem{sondhi}
T.~Hansson, V.~Oganesyan, S.~Sondhi, Superconductors are topologically ordered,
  Annals of Physics 313~(2) (2004) 497 -- 538.
\newblock \href {https://doi.org/10.1016/j.aop.2004.05.006}
  {\path{doi:10.1016/j.aop.2004.05.006}}.

\bibitem{lucile}
L.~Savary, L.~Balents,
  \href{http://stacks.iop.org/0034-4885/80/i=1/a=016502}{Quantum spin liquids:
  a review}, Reports on Progress in Physics 80~(1) (2017) 016502.
\newline\urlprefix\url{http://stacks.iop.org/0034-4885/80/i=1/a=016502}

\bibitem{moroz}
S.~Moroz, A.~Prem, V.~Gurarie, L.~Radzihovsky, Topological order, symmetry, and
  hall response of two-dimensional spin-singlet superconductors, Phys. Rev. B
  95 (2017) 014508.
\newblock \href {https://doi.org/10.1103/PhysRevB.95.014508}
  {\path{doi:10.1103/PhysRevB.95.014508}}.

\bibitem{kalb}
M.~Kalb, P.~Ramond, Classical direct interstring action, Phys. Rev. D 9 (1974)
  2273--2284.
\newblock \href {https://doi.org/10.1103/PhysRevD.9.2273}
  {\path{doi:10.1103/PhysRevD.9.2273}}.

\bibitem{kapustin}
A.~Kapustin, R.~Thorngren, Higher symmetry and gapped phases of gauge theories
  (Sep).
\newblock \href {http://arxiv.org/abs/1309.4721} {\path{arXiv:1309.4721}}.

\bibitem{vasiliev}
M.~Vasiliev, Higher-spin gauge theories in four, three, and two dimensions,
  International Journal of Modern Physics D 05~(06) (1996) 763--797.
\newblock \href {https://doi.org/10.1142/S0218271896000473}
  {\path{doi:10.1142/S0218271896000473}}.

\bibitem{cenke}
C.~Xu, Gapless bosonic excitation without symmetry breaking: An algebraic spin
  liquid with soft gravitons, Phys. Rev. B 74 (2006) 224433.
\newblock \href {https://doi.org/10.1103/PhysRevB.74.224433}
  {\path{doi:10.1103/PhysRevB.74.224433}}.

\bibitem{wengu}
Z.-C. Gu, X.-G. Wen, Emergence of helicity ±2 modes (gravitons) from qubit
  models, Nuclear Physics B 863 (2012) 90--129.
\newblock \href {https://doi.org/10.1016/j.nuclphysb.2012.05.010}
  {\path{doi:10.1016/j.nuclphysb.2012.05.010}}.

\bibitem{horava}
C.~Xu, P.~Ho\ifmmode~\check{r}\else \v{r}\fi{}ava, Emergent gravity at a
  lifshitz point from a bose liquid on the lattice, Phys. Rev. D 81 (2010)
  104033.
\newblock \href {https://doi.org/10.1103/PhysRevD.81.104033}
  {\path{doi:10.1103/PhysRevD.81.104033}}.

\bibitem{rasmussen}
A.~{Rasmussen}, Y.-Z. {You}, C.~{Xu}, {Stable Gapless Bose Liquid Phases
  without any Symmetry} (Jan.).
\newblock \href {http://arxiv.org/abs/1601.08235} {\path{arXiv:1601.08235}}.

\bibitem{sub}
M.~Pretko, Subdimensional particle structure of higher rank $u(1)$ spin
  liquids, Phys. Rev. B 95 (2017) 115139.
\newblock \href {https://doi.org/10.1103/PhysRevB.95.115139}
  {\path{doi:10.1103/PhysRevB.95.115139}}.

\bibitem{genem}
M.~Pretko, Generalized electromagnetism of subdimensional particles: A spin
  liquid story, Phys. Rev. B 96 (2017) 035119.
\newblock \href {https://doi.org/10.1103/PhysRevB.96.035119}
  {\path{doi:10.1103/PhysRevB.96.035119}}.

\bibitem{higgs1}
H.~{Ma}, M.~{Hermele}, X.~{Chen}, {Fracton topological order from Higgs and
  partial confinement mechanisms of rank-two gauge theory} (Feb. 2018).
\newblock \href {http://arxiv.org/abs/1802.10108} {\path{arXiv:1802.10108}}.

\bibitem{higgs2}
D.~Bulmash, M.~Barkeshli, Higgs mechanism in higher-rank symmetric u(1) gauge
  theories, Phys. Rev. B 97 (2018) 235112.
\newblock \href {https://doi.org/10.1103/PhysRevB.97.235112}
  {\path{doi:10.1103/PhysRevB.97.235112}}.

\bibitem{haah}
J.~Haah, Local stabilizer codes in three dimensions without string logical
  operators, Phys. Rev. A 83 (2011) 042330.
\newblock \href {https://doi.org/10.1103/PhysRevA.83.042330}
  {\path{doi:10.1103/PhysRevA.83.042330}}.

\bibitem{chamon}
C.~Chamon, Quantum glassiness in strongly correlated clean systems: An example
  of topological overprotection, Phys. Rev. Lett. 94 (2005) 040402.
\newblock \href {https://doi.org/10.1103/PhysRevLett.94.040402}
  {\path{doi:10.1103/PhysRevLett.94.040402}}.

\bibitem{castelnovo}
C.~Castelnovo, C.~Chamon, Topological quantum glassiness, Philosophical
  Magazine 92~(1-3) (2012) 304--323.
\newblock \href {https://doi.org/10.1080/14786435.2011.609152}
  {\path{doi:10.1080/14786435.2011.609152}}.

\bibitem{bravyi}
S.~Bravyi, B.~Leemhuis, B.~M. Terhal, Topological order in an exactly solvable
  3d spin model, Annals of Physics 326~(4) (2011) 839 -- 866.
\newblock \href {https://doi.org/10.1016/j.aop.2010.11.002}
  {\path{doi:10.1016/j.aop.2010.11.002}}.

\bibitem{haah2}
S.~Bravyi, J.~Haah, Quantum self-correction in the 3d cubic code model, Phys.
  Rev. Lett. 111 (2013) 200501.
\newblock \href {https://doi.org/10.1103/PhysRevLett.111.200501}
  {\path{doi:10.1103/PhysRevLett.111.200501}}.

\bibitem{yoshida}
B.~Yoshida, Exotic topological order in fractal spin liquids, Phys. Rev. B 88
  (2013) 125122.
\newblock \href {https://doi.org/10.1103/PhysRevB.88.125122}
  {\path{doi:10.1103/PhysRevB.88.125122}}.

\bibitem{fracton1}
S.~Vijay, J.~Haah, L.~Fu, A new kind of topological quantum order: A
  dimensional hierarchy of quasiparticles built from stationary excitations,
  Phys. Rev. B 92 (2015) 235136.
\newblock \href {https://doi.org/10.1103/PhysRevB.92.235136}
  {\path{doi:10.1103/PhysRevB.92.235136}}.

\bibitem{fracton2}
S.~Vijay, J.~Haah, L.~Fu, Fracton topological order, generalized lattice gauge
  theory, and duality, Phys. Rev. B 94 (2016) 235157.
\newblock \href {https://doi.org/10.1103/PhysRevB.94.235157}
  {\path{doi:10.1103/PhysRevB.94.235157}}.

\bibitem{williamson}
D.~J. Williamson, Fractal symmetries: Ungauging the cubic code, Phys. Rev. B 94
  (2016) 155128.
\newblock \href {https://doi.org/10.1103/PhysRevB.94.155128}
  {\path{doi:10.1103/PhysRevB.94.155128}}.

\bibitem{DevakulWilliamson}
T.~Devakul, D.~J. Williamson, Universal quantum computation using fractal
  symmetry-protected cluster phases (Jun 2018).
\newblock \href {http://arxiv.org/abs/1806.04663} {\path{arXiv:1806.04663}}.

\bibitem{prem}
A.~Prem, J.~Haah, R.~Nandkishore, Glassy quantum dynamics in translation
  invariant fracton models, Phys. Rev. B 95 (2017) 155133.
\newblock \href {https://doi.org/10.1103/PhysRevB.95.155133}
  {\path{doi:10.1103/PhysRevB.95.155133}}.

\bibitem{han}
H.~Ma, E.~Lake, X.~Chen, M.~Hermele, Fracton topological order via coupled
  layers, Phys. Rev. B 95 (2017) 245126.
\newblock \href {https://doi.org/10.1103/PhysRevB.95.245126}
  {\path{doi:10.1103/PhysRevB.95.245126}}.

\bibitem{sagar}
S.~{Vijay}, {Isotropic Layer Construction and Phase Diagram for Fracton
  Topological Phases} (Jan.).
\newblock \href {http://arxiv.org/abs/1701.00762} {\path{arXiv:1701.00762}}.

\bibitem{mach}
M.~Pretko, Emergent gravity of fractons: Mach's principle revisited, Phys. Rev.
  D 96 (2017) 024051.
\newblock \href {https://doi.org/10.1103/PhysRevD.96.024051}
  {\path{doi:10.1103/PhysRevD.96.024051}}.

\bibitem{hsieh}
T.~H. Hsieh, G.~B. Hal\'asz, Fractons from partons, Phys. Rev. B 96 (2017)
  165105.
\newblock \href {https://doi.org/10.1103/PhysRevB.96.165105}
  {\path{doi:10.1103/PhysRevB.96.165105}}.

\bibitem{slagle1}
K.~Slagle, Y.~B. Kim, Fracton topological order from nearest-neighbor two-spin
  interactions and dualities, Phys. Rev. B 96 (2017) 165106.
\newblock \href {https://doi.org/10.1103/PhysRevB.96.165106}
  {\path{doi:10.1103/PhysRevB.96.165106}}.

\bibitem{screening}
M.~Pretko, Finite-temperature screening of $u$(1) fractons, Phys. Rev. B 96
  (2017) 115102.
\newblock \href {https://doi.org/10.1103/PhysRevB.96.115102}
  {\path{doi:10.1103/PhysRevB.96.115102}}.

\bibitem{nonabelian}
S.~{Vijay}, L.~{Fu}, {A Generalization of Non-Abelian Anyons in Three
  Dimensions} (Jun.).
\newblock \href {http://arxiv.org/abs/1706.07070} {\path{arXiv:1706.07070}}.

\bibitem{decipher}
B.~Shi, Y.-M. Lu, Deciphering the nonlocal entanglement entropy of fracton
  topological orders, Phys. Rev. B 97 (2018) 144106.
\newblock \href {https://doi.org/10.1103/PhysRevB.97.144106}
  {\path{doi:10.1103/PhysRevB.97.144106}}.

\bibitem{balents}
G.~B. Hal\'asz, T.~H. Hsieh, L.~Balents, Fracton topological phases from
  strongly coupled spin chains, Phys. Rev. Lett. 119 (2017) 257202.
\newblock \href {https://doi.org/10.1103/PhysRevLett.119.257202}
  {\path{doi:10.1103/PhysRevLett.119.257202}}.

\bibitem{slagle2}
K.~Slagle, Y.~B. Kim, Quantum field theory of x-cube fracton topological order
  and robust degeneracy from geometry, Phys. Rev. B 96 (2017) 195139.
\newblock \href {https://doi.org/10.1103/PhysRevB.96.195139}
  {\path{doi:10.1103/PhysRevB.96.195139}}.

\bibitem{chiral}
M.~Pretko, Higher-spin witten effect and two-dimensional fracton phases, Phys.
  Rev. B 96 (2017) 125151.
\newblock \href {https://doi.org/10.1103/PhysRevB.96.125151}
  {\path{doi:10.1103/PhysRevB.96.125151}}.

\bibitem{prem2}
A.~Prem, M.~Pretko, R.~M. Nandkishore, Emergent phases of fractonic matter,
  Phys. Rev. B 97 (2018) 085116.
\newblock \href {https://doi.org/10.1103/PhysRevB.97.085116}
  {\path{doi:10.1103/PhysRevB.97.085116}}.

\bibitem{regnault}
O.~Petrova, N.~Regnault, Simple anisotropic three-dimensional quantum spin
  liquid with fractonlike topological order, Phys. Rev. B 96 (2017) 224429.
\newblock \href {https://doi.org/10.1103/PhysRevB.96.224429}
  {\path{doi:10.1103/PhysRevB.96.224429}}.

\bibitem{valbert}
V.~V. Albert, S.~Pascazio, M.~H. Devoret,
  \href{http://stacks.iop.org/1751-8121/50/i=50/a=504002}{General phase spaces:
  from discrete variables to rotor and continuum limits}, Journal of Physics A:
  Mathematical and Theoretical 50~(50) (2017) 504002.
\newline\urlprefix\url{http://stacks.iop.org/1751-8121/50/i=50/a=504002}

\bibitem{devakul}
T.~Devakul, S.~A. Parameswaran, S.~L. Sondhi, Correlation function diagnostics
  for type-i fracton phases, Phys. Rev. B 97 (2018) 041110.
\newblock \href {https://doi.org/10.1103/PhysRevB.97.041110}
  {\path{doi:10.1103/PhysRevB.97.041110}}.

\bibitem{regnault2}
H.~He, Y.~Zheng, B.~A. Bernevig, N.~Regnault, Entanglement entropy from tensor
  network states for stabilizer codes, Phys. Rev. B 97 (2018) 125102.
\newblock \href {https://doi.org/10.1103/PhysRevB.97.125102}
  {\path{doi:10.1103/PhysRevB.97.125102}}.

\bibitem{han2}
H.~Ma, A.~T. Schmitz, S.~A. Parameswaran, M.~Hermele, R.~M. Nandkishore,
  Topological entanglement entropy of fracton stabilizer codes, Phys. Rev. B 97
  (2018) 125101.
\newblock \href {https://doi.org/10.1103/PhysRevB.97.125101}
  {\path{doi:10.1103/PhysRevB.97.125101}}.

\bibitem{albert}
A.~T. Schmitz, H.~Ma, R.~M. Nandkishore, S.~A. Parameswaran, Recoverable
  information and emergent conservation laws in fracton stabilizer codes, Phys.
  Rev. B 97 (2018) 134426.
\newblock \href {https://doi.org/10.1103/PhysRevB.97.134426}
  {\path{doi:10.1103/PhysRevB.97.134426}}.

\bibitem{leomichael}
M.~Pretko, L.~Radzihovsky, Fracton-elasticity duality, Phys. Rev. Lett. 120
  (2018) 195301.
\newblock \href {https://doi.org/10.1103/PhysRevLett.120.195301}
  {\path{doi:10.1103/PhysRevLett.120.195301}}.

\bibitem{gromov}
A.~{Gromov}, {Fractional Topological Elasticity and Fracton Order} (Dec.).
\newblock \href {http://arxiv.org/abs/1712.06600} {\path{arXiv:1712.06600}}.

\bibitem{shirley}
W.~{Shirley}, K.~{Slagle}, Z.~{Wang}, X.~{Chen}, {Fracton Models on General
  Three-Dimensional Manifolds} (Dec.).
\newblock \href {http://arxiv.org/abs/1712.05892} {\path{arXiv:1712.05892}}.

\bibitem{foliatedEntanglement}
W.~Shirley, K.~Slagle, X.~Chen, Universal entanglement signatures of foliated
  fracton phases (Mar 2018).
\newblock \href {http://arxiv.org/abs/1803.10426} {\path{arXiv:1803.10426}}.

\bibitem{foliatedExcitations}
W.~Shirley, K.~Slagle, X.~Chen, Fractional excitations in foliated fracton
  phases (Jun 2018).
\newblock \href {http://arxiv.org/abs/1806.08625} {\path{arXiv:1806.08625}}.

\bibitem{shirleyCheckerboard}
W.~Shirley, K.~Slagle, X.~Chen, Foliated fracton order in the checkerboard
  model (Jun 2018).
\newblock \href {http://arxiv.org/abs/1806.08633} {\path{arXiv:1806.08633}}.

\bibitem{shirleyGauging}
W.~Shirley, K.~Slagle, X.~Chen, Foliated fracton order from gauging subsystem
  symmetries (Jun 2018).
\newblock \href {http://arxiv.org/abs/1806.08679} {\path{arXiv:1806.08679}}.

\bibitem{slagle3}
K.~Slagle, Y.~B. Kim, X-cube model on generic lattices: Fracton phases and
  geometric order, Phys. Rev. B 97 (2018) 165106.
\newblock \href {https://doi.org/10.1103/PhysRevB.97.165106}
  {\path{doi:10.1103/PhysRevB.97.165106}}.

\bibitem{pai}
S.~{Pai}, M.~{Pretko}, {Fractonic line excitations : an inroad from 3d
  elasticity theory} (Apr.).
\newblock \href {http://arxiv.org/abs/1804.01536} {\path{arXiv:1804.01536}}.

\bibitem{yizhi1}
Y.~{You}, T.~{Devakul}, F.~J. {Burnell}, S.~L. {Sondhi}, {Subsystem symmetry
  protected topological order} (Mar.).
\newblock \href {http://arxiv.org/abs/1803.02369} {\path{arXiv:1803.02369}}.

\bibitem{cagenet}
A.~Prem, S.-J. Huang, H.~Song, M.~Hermele, Cage-net fracton models (Jun 2018).
\newblock \href {http://arxiv.org/abs/1806.04687} {\path{arXiv:1806.04687}}.

\bibitem{twisted}
H.~{Song}, A.~{Prem}, S.-J. {Huang}, M.~A. {Martin-Delgado}, {Twisted Fracton
  Models in Three Dimensions} (May).
\newblock \href {http://arxiv.org/abs/1805.06899} {\path{arXiv:1805.06899}}.

\bibitem{deconfined}
H.~{Ma}, M.~{Pretko}, {Higher Rank Deconfined Quantum Criticality and the
  Exciton Bose Condensate} (Mar. 2018).
\newblock \href {http://arxiv.org/abs/1803.04980} {\path{arXiv:1803.04980}}.

\bibitem{ungauging}
A.~{Kubica}, B.~{Yoshida}, {Ungauging quantum error-correcting codes} (May
  2018).
\newblock \href {http://arxiv.org/abs/1805.01836} {\path{arXiv:1805.01836}}.

\bibitem{fractalsym}
T.~{Devakul}, Y.~{You}, F.~J. {Burnell}, S.~L. {Sondhi}, {Fractal Symmetric
  Phases of Matter} (May 2018).
\newblock \href {http://arxiv.org/abs/1805.04097} {\path{arXiv:1805.04097}}.

\bibitem{symfrac}
Y.~{You}, T.~{Devakul}, F.~J. {Burnell}, S.~L. {Sondhi}, {Symmetric Fracton
  Matter: Twisted and Enriched} (May 2018).
\newblock \href {http://arxiv.org/abs/1805.09800} {\path{arXiv:1805.09800}}.

\bibitem{BulmashFractal}
D.~Bulmash, M.~Barkeshli, Generalized $u(1)$ gauge field theories and fractal
  dynamics (Jun 2018).
\newblock \href {http://arxiv.org/abs/1806.01855} {\path{arXiv:1806.01855}}.

\bibitem{pinchpoints}
A.~Prem, S.~Vijay, Y.-Z. Chou, M.~Pretko, R.~M. Nandkishore, Pinch point
  singularities of tensor spin liquids (Jun 2018).
\newblock \href {http://arxiv.org/abs/1806.04148} {\path{arXiv:1806.04148}}.

\bibitem{fractonreview}
R.~M. {Nandkishore}, M.~{Hermele}, {Fractons} (Mar.).
\newblock \href {http://arxiv.org/abs/1803.11196} {\path{arXiv:1803.11196}}.

\bibitem{asymptotic1}
W.~De~Roeck, F.~Huveneers, Asymptotic quantum many-body localization from
  thermal disorder, Communications in Mathematical Physics 332~(3) (2014)
  1017--1082.
\newblock \href {https://doi.org/10.1007/s00220-014-2116-8}
  {\path{doi:10.1007/s00220-014-2116-8}}.

\bibitem{asymptotic2}
W.~De~Roeck, F.~m.~c. Huveneers, Scenario for delocalization in
  translation-invariant systems, Phys. Rev. B 90 (2014) 165137.
\newblock \href {https://doi.org/10.1103/PhysRevB.90.165137}
  {\path{doi:10.1103/PhysRevB.90.165137}}.

\bibitem{Carroll}
S.~M. Carroll, Lecture notes on general relativity\href
  {http://arxiv.org/abs/gr-qc/9712019} {\path{arXiv:gr-qc/9712019}}.

\bibitem{Nakahara}
M.~Nakahara, Geometry, Topology and Physics, Second Edition, 2nd Edition,
  Institute of Physics, 2003.

\bibitem{Weinberg}
S.~Weinberg, Gravitation and Cosmology: Principles and Applications of the
  General Theory of Relativity, John Wiley \& Sons, Inc., 1972.

\bibitem{notebook}
See the ancillary file
  `\href{https://arxiv.org/src/1807.00827v2/anc/curvedFractons.nb}{curvedFractons.nb}'
  attached to our arXiv submission for our Mathematica notebook used to check
  gauge invariance.

\bibitem{haahU1}
J.~Haah, Two generalizations of the cubic code model.\,\,Talk at KITP, Oct. 13
  2017.

\end{thebibliography}

%---------------------------------------
%---------------------------------------
\end{document}